# BERT-PIN: A BERT-based Framework for Recovering Missing Data Segments in Time-series Load Profiles

Yi Hu, *Student Member, IEEE*, Kai Ye, *Student Member, IEEE*, Hyeonjin Kim, *Student Member, IEEE*, and* Ning Lu, *Fellow, IEEE*

*Abstract*—**Inspired by the success of the Transformer model in natural language processing and computer vision, this paper introduces BERT-PIN, a Bidirectional Encoder Representations from Transformers (BERT) powered Profile Inpainting Network. BERT−PIN recovers multiple missing data segments (MDSs) using load and temperature time-series profiles as inputs. To adopt a standard Transformer model structure for profile inpainting, we segment the load and temperature profiles into line segments, treating each segment as a word and the entire profile as a sentence. We incorporate a top candidates selection process in BERT-PIN, enabling it to produce a sequence of probability distributions, based on which users can generate multiple plausible imputed data sets, each reflecting different confidence levels. We develop and evaluate BERT-PIN using real-world dataset for two applications: multiple MDSs recovery and demand response baseline estimation. Simulation results show that BERT−PIN outperforms the existing methods in accuracy while is capable of restoring multiple MDSs within a longer window. BERT-PIN, served as a pre-trained model, can be fine-tuned for conducting many downstream tasks, such as classification and super resolution.**

## I. INTRODUCTION

MISSING data restoration plays a critical role in power system analysis. In [1], we conducted a comprehensive literature review to assess the performance of both model-based and data-driven load profile inpainting methods [2]-[19] for the restoration of missing data. Additionally, in [1], we demonstrated that our GAN-based approach outperforms existing methods when it comes to restoring missing data segments (MDSs) of fixed length. This GAN-based method uses specialized attention patterns to capture and model complex dependencies and structures within the data.

Since 2017, the Transformer model [20] has garnered remarkable success in the field of natural language processing (NLP) by excelling at handling sequential data and capturing extensive long-range dependencies through simultaneous attention to all positions in a sequence. The introduction of self-attention enables models to discern the importance of individual elements within input data. This capability enables the Transformer model to grasp intricate dependencies, relationships, and contextual information, resulting in improved performance in other domains, such as speech recognition [21], and computer vision (CV) [22].

Inspired by the success of the Bidirectional Encoder Representations from Transformers (BERT) model, we introduce BERT-PIN, a BERT-based Profile Inpainting Network (BERT-PIN) as a versatile framework for restoring MDSs within a long time-series load profile. Note that for a comprehensive overview of load profile inpainting methodologies, please refer to [1]. In this paper, we will focus on reviewing BERT-based methods that are more relevant to our proposed approach.

In CV domain, the authors proposed Vision Transformer (ViT) [22], which involves dividing images into patches and converting them into a sequence of linear embeddings. These embeddings are then used as inputs to a Transformer model. This process empowers the Transformer to directly process and comprehend visual information through self-attention mechanisms, achieving state-of-the-art performance in tasks such as image classification and object detection [23]-[26].

Inspired by ViT, we propose the BERT-based Profile Inpainting Network (BERT-PIN), a versatile framework tailored to restoring multiple absent data segments. To adopt a standard Transformer model structure for profile inpainting, we segment the load and temperature profiles into line segments, treating each segment as a word, the daily profile as a sentence, and the weekly/monthly load profile as a paragraph consisting multiple sentences. Additionally, we incorporate a top candidates selection process in BERT-PIN, enabling it to produce a sequence of probability distributions, based on which users can generate multiple plausible imputed data sets, each reflecting different confidence levels.

As shown in Fig. 1, MDSs (depicted in green) can be treated as missing words in sentences, making the restoration process equivalent to recovering missing words in sentences or paragraphs. One of the advantages of employing BERT is its

This material is based upon work supported by the U.S. Department of Energy's Office of Energy Efficiency and Renewable Energy (EERE) under the Solar Energy Technologies Office. Award Number: DE-EE0008770.

Yi Hu, Kai Ye, Hyeonjin Kim, and Ning Lu are with the Electrical & Computer Engineering Department, Future Renewable Energy Delivery and Management (FREEDM) Systems Center, North Carolina State University, Raleigh, NC 27606 USA. (yhu28@ncsu.edu, kye3@ncsu.edu, hkim66@ncsu.edu, nlu2@ncsu.edu).

The source code is available at https://github.com/hughwln/BERT-PIN_public.

ability to generate multiple alternatives for a missing word, each accompanied by an associated confidence level.

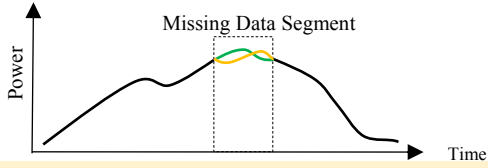

Fig. 1. An illustration of the load profile inpainting problem.

In the word filling task illustrated in Fig. 1, if someone is proficient in speaking French, "France" may emerge as the most probable response. Nevertheless, "Quebec" and "Cameroon" also stand as viable alternatives, as people from these regions also use French as an official language. Likewise, when restoring an MDS, multiple patching options are available, each with a comparable likelihood of being the best match. Nevertheless, the majority of existing inpainting techniques only provide a single viable solution.

Our *primary* contribution is, to the best of our knowledge, the introduction of a BERT-based approach to load profile inpainting. In evaluations conducted on a real-world dataset consisting of 3 years of 15-minute smart meter data from 8000 users, BERT-PIN outperforms the GAN-based method from [1] by approximately 10% in overall load profile inpainting. It's worth noting that the authors of [1] conducted a thorough comparative analysis with other established methods, demonstrating its superior performance when compared to these alternatives. Our *second* contribution is that, unlike existing methods, BERT-PIN can produce multiple data restoration candidates with varying levels of confidence. This feature is particularly valuable when it is necessary to explore all potential options to ensure the robustness of an algorithm. *Lastly*, in addition to its accuracy and the capability to generate an ensemble of restoration options, BERT-PIN also facilitates the use of flexible masking strategies, enabling restoring multiple missing data segments (MDSs) within very long-time windows. This flexibility allows BERT-PIN to be used as a pre-trained model for many downstream tasks, such as load profile disaggregation [27] and super resolution [28].

The rest of the paper is organized as follows. Section II introduces the methodology, Section III introduces the simulation results in different cases, and Section IV concludes the paper.

## II. METHODOLOGY

In this section, the load profile inpainting problem formulation, the BERT-PIN model architecture, and the performance evaluation metrics are introduced.

### A. BERT-PIN Problem Formulation

Define $\boldsymbol{X}_m$ as a MDS (the green segment in Fig. 2) in a time series load profile, $\boldsymbol{X} = [x_1, x_2, ..., x_N]$, where $N$ denotes the length of the time series.

As shown in Fig. 2, if $\boldsymbol{X}$ contains $N_m$ MDSs, i.e., $[\boldsymbol{X}_{m,1}, \boldsymbol{X}_{m,2}, ...\boldsymbol{X}_{m,i}, ..., \boldsymbol{X}_{m,N_m}]$, the objective of the inpainting problem is to find a set of model parameters, $\theta$, to recover all MDSs using the non-missing data $\widetilde{\boldsymbol{X}}$ in $\boldsymbol{X}$ and the corresponding ambient temperature profile, $\boldsymbol{T} = [x_1, x_2, ..., x_N]$ as inputs. So, the problem can be formulated as

$$[\widehat{\boldsymbol{X}}_{m,1}, \widehat{\boldsymbol{X}}_{m,2}, ..., \widehat{\boldsymbol{X}}_{m,i}, ..., \widehat{\boldsymbol{X}}_{m,N_m}] = f_\theta(T, \widetilde{\boldsymbol{X}}) \quad (1)$$

where $\widehat{\boldsymbol{X}}_{m,i}$ is the $i$th recovered MDS.

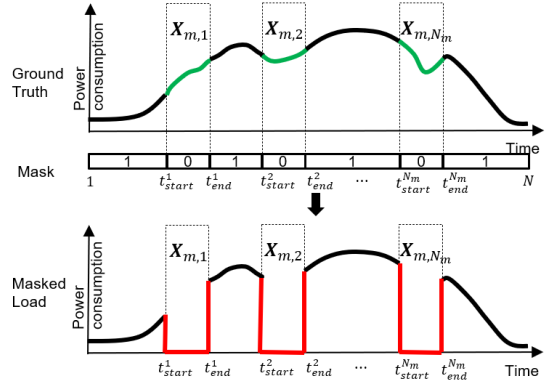

Fig. 2. Illustration of missing data segments and masking methods.

Although many machine-learning based methods can be applied for restoring missing data, we chose the BERT-based model for three reasons. *First*, BERT is a bi-directional model, so it is well-suited for capturing contextual information of the MDSs within a time-series load profile. *Second*, BERT leverages self-attention mechanisms to effectively capture long-range dependencies for enhancing its performance in processing sequential data. *Third*, BERT architecture is highly flexible when accommodating multi-modality inputs, such as combining the temperature profile with the load profile.

As shown in Fig. 3, the proposed BERT-PIN model contains three key processes: input data adaptation, BERT model for recovering MDSs, and top candidates selection.

Given the proven effectiveness of the BERT model structure in handling NLP tasks, we do not intend to change the original BERT model architecture. However, as the BERT model is initially designed for processing NLP problems, its inputs are a sequence of word tokens. Hence, in the *Input Data Adaptation* process, we first align the load profile with its corresponding temperature profile. Then, we divide the two aligned time-series profiles into segments to generate the load and temperature embedding, respectively. As shown in Figs. 1 and3, when each of these segments is analogous to a missing word in a sentence, the task of restoring missing data becomes equivalent to recovering missing words within sentences or paragraphs.

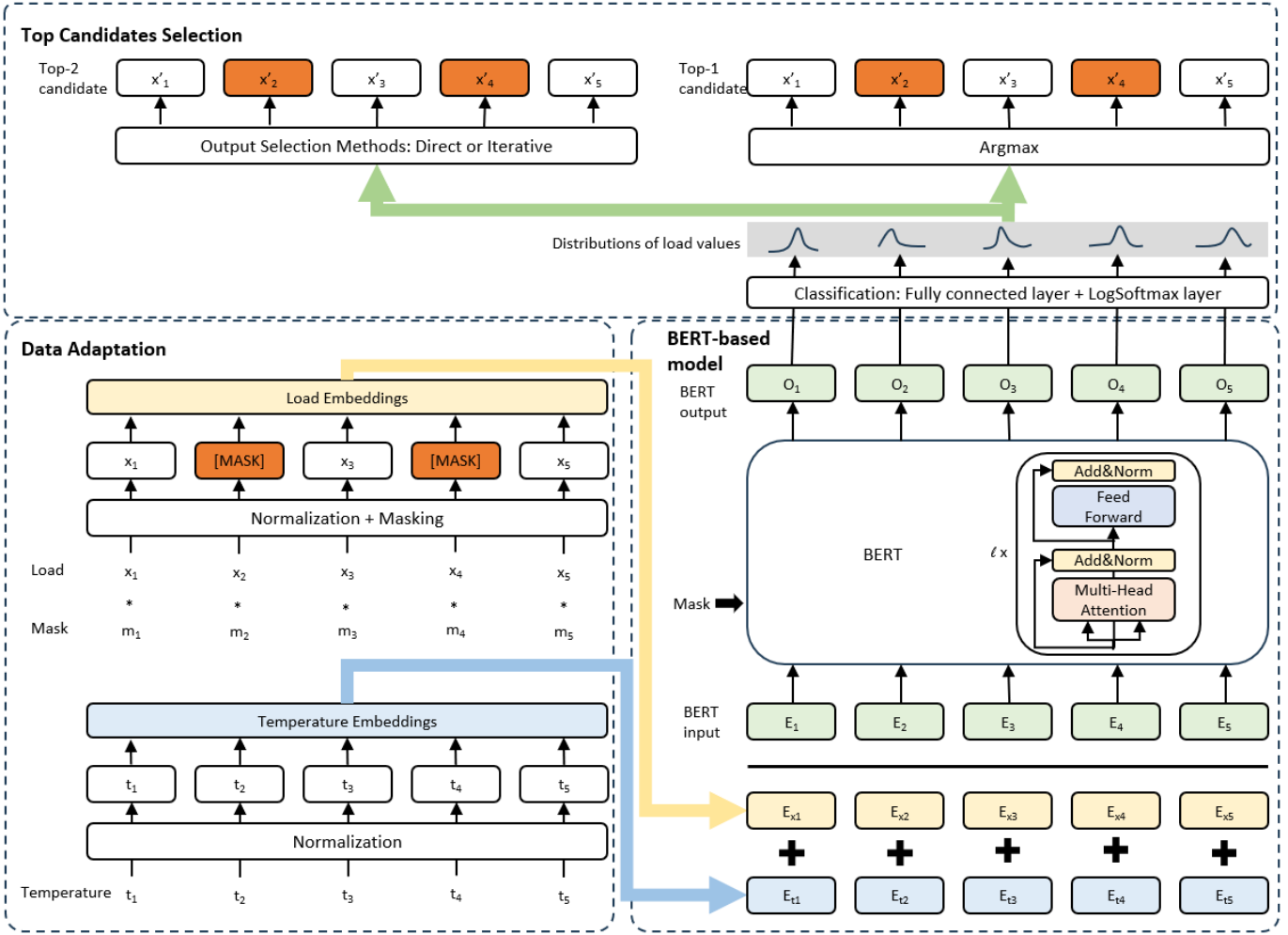

Fig. 3. An overview of the overall modeling framework. The illustration of the BERT was inspired by [20]. It's an example of sequence length $N = 5$.

In the *Top Candidates Selection* process, rather than selecting a single candidate based on the highest likelihood, our goal is to generate multiple candidates that fall within specified confidence levels. This capability allows BERT-PIN to be used for producing an ensemble of patching options for an MDS. This function becomes particularly crucial when dealing with MDSs where several candidates exhibit similar probabilities of being the potential outcomes.

We consider those two augmentations to the existing BERT model structure as our contributions.

### B. Training Data Preparation

#### 1) Preparation of Ground Truth Load Profiles

To prepare the training data, we gathered 15-minute interval smart meter data spanning three-year (2018-2020) from 8000 residential households in North Carolina. Let $P_i$ be the load profile for the $i^{th}$ user containing $N$ data points. To begin, we select a starting time ($t_{start}$) and calculate the end time ($t_{end}$) of the time series by $t_{end} = t_{start} + N - 1$. Next, we randomly draw 2000 load profiles ($P$) from the pool of 8000 load profiles and aggregate them into one load profile. We normalize the aggregated load profile by its peak power ($P_{MAX}$) to create the ground truth load profile ($X$). This ensures that $X$ falls within the range of [0, 1]. This process can be described as

$$X = \frac{1}{P_{MAX}} \sum_{i=1}^{2000} P_i \quad (t_{start} : t_{end}) \quad (2)$$

$$P_{MAX} = \max \sum_{i=1}^{2000} P_i \quad (0 : N) \quad (3)$$

#### 2) Preparation of Load Profiles with Missing Data

To generate the time series data with $N_m$ MDSs, we create a mask vector $\boldsymbol{M}$ for each $\boldsymbol{X}$, so that

$$M = [m_i \text{ for } i = 1 : N], m_i = \begin{cases} 0, & \text{missing data} \\ 1, & \text{otherwise} \end{cases} \quad (4)$$

Then, the masked load profile, $\boldsymbol{X^m}$, can be represented as

$$\boldsymbol{X^m} = \boldsymbol{X} \cdot \boldsymbol{M} \quad (5)$$

Note that we set all missing data segments to be 0 kW because the power range of the aggregated load profiles typically falls within [210, 1751] kW. We want to underscore that assigning missing data a value well outside the allowable data range is crucial to ensure clear differentiation from normal values.

In this paper, we apply two masking strategies: central-masking and peak-masking. With central masking, we place the MDS in the middle of the sequence. Conversely, with peak-masking, we position the MDS during periods of highest energy consumption.

Although the number of MDSs may vary, we set the length of the MDS to be fixed at 4 hours to reduce the training time and computing cost. This choice is made also because the majority of MDS durations are shorter than 4 hours. For example, when examining the 15-minute smart meter data collected from 8000 users over a three-year period, we find that approximately 70% of the MDSs have durations of less than 4 hours.

### 3) Input Data Adaptation Layer

To align the BERT-PIN inputs with the BERT required format, we map the values in $X^m$ to integers between 0 to 200. Note that we chose 200 because it strikes a balance between the model's size and resolution. In our dataset, the power range is [210, 1751] kW, as shown in Fig. 4. So, the mapping provides a resolution of 8.755 kW.

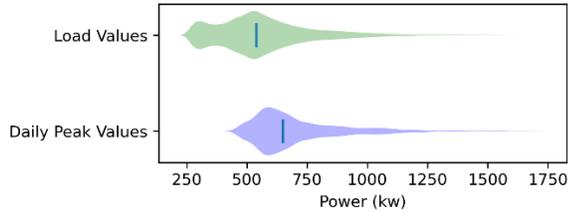

Fig. 4. Distribution of the aggregated load values and daily peak values.

The mapped load profile is embedded into a $N \times 200$ matrix, represented by the load embeddings (yellow boxes) in Fig. 3. This data adaptation process allows the model to generate a probability distribution for each data point, making it possible to generate an ensemble of candidates. This transforms the missing data restoration problem, which is usually a regression problem, into a classification problem.

To address the influence of temperature on load [29][30], we include the normalized ambient temperature profile data, designated as $T$, as an additional modality input to assist in the recovery of missing load data. $T$ is subjected to normalization based on the highest and lowest annual temperatures, ensuring that the normalized temperature values also range from 0 to 1. Then, $T$ is rescaled to the [0, 200] range using the same approach employed for load embedding.

Lastly, we combine $X^m$ and $T$ embeddings together by element-wise addition to obtain the final input matrix, the dimension of which is $N \times 200$. As an illustration, we show the input data adaptation process when $N = 5$ in Fig. 3.

### C. BERT Model

The combined embeddings of load and temperature can be directly fed into the BERT model using its original model architecture introduced in [31]. Represent a data sequence as $D = \{\{k_1, v_1\}, ... \{k_N, v_N\}\}$, where $k$ and $v$ are $N$ tuples of keys and values, respectively. For a query $q$, the attention over $D$ in the BERT model is formulated as

$$Attention(q, D) = \sum_{i=1}^{N} \alpha(q, k_i) \, v_i \quad (6)$$

where $\alpha(q, k_i) \in \mathbb{R}$ ($i = 1, ..., N$) are scalar values calculated through the dot product of the query vector and each key vector. These weights, referred to as attention scores, are normalized to ensure a sum of 1. This calculation is performed for every element in the input sequence. The output vectors are combined to form the final output of the model.

When processing sequential data, using self-attention can selectively attend to different parts of the input sequence based on their relevance to the given query vector. Thus, the BERT model can effectively capture long-range dependencies to form a context for the sequence. This significantly facilitates the recovery of the missing data because the context reflects the relevance of all known data points from all modalities (e.g., load and temperature) with the missing data points in those time-series profiles.

In the training, we use $\hat{X}^1$ to calculate the cross-entropy losses as

$$CrossEntropy = -\frac{1}{N}\sum_{o=1}^{N}\sum_{c=1}^{C} x_{o,c} \log(d_{o,c}) \quad (7)$$

where $C$ is the number of classes, $x_{o,c}$ is the truth label denoting the power consumption value for observation $o$, and $d_{o,c}$ is the predicted probability observation $o$ belonging to class $c$. $N$ is the length of the sequence.

Note that the objective of BERT-PIN is to effectively restore the missing segments in a load profile, so we need to train the network to place more focus on restoring the MDSs. Thus, we construct the loss function as

$$Loss = (1-\lambda) * CrossEntropy(X, \hat{X}^1)$$
$$+ \lambda * CrossEntropy(X_m, \hat{X}_m^1) \quad (8)$$

where $\lambda$ is a hyper parameter for balancing between the global and local losses. Thus, $(1-\lambda) * CrossEntropy(X, \hat{X}^1)$ represents the global loss for assessing how well the whole load profile can be restored, and $\lambda * CrossEntropy(X_m, \hat{X}_m^1)$ represents the local loss for assessing how well the MDSs are restored.

### D. Top Candidates Selection Layer

As illustrated in Fig. 3, we add a *Top Candidates Selection* process. Define the output of the BERT model, $O$, as:

$$O = BERT(X^m, T) \quad (9)$$

where $O$ is an $N \times 200$ matrix.

We feed $O$ into a classification layer, which comprises a fully connected layer followed by a SoftMax layer, to obtain an $N \times 200$ probability distribution matrix, $D$. Note that the $i$th column of $D$ represents the probability distribution function (PDF) of the value of the $i$th data point falling within the range of 1 to 200.

$$D = Classification(O) \quad (10)$$

Using $D$, we can generate an ensemble of curves for patching an MDS rather than outputting just a single curve with the highest likelihood.

The conventional method for restoring the MDS from $D$ is to use an *argmax* layer as illustrated as the orange boxes in Fig. 3 by

$$\hat{X}^1 = argmax(D) \quad (11)$$

where $\hat{X}^1$ is considered as the *top-1 candidate*.

Nevertheless, it is often necessary to explore the top-2 or even the top-3 candidates as potential patching options to enhance the inclusion. As depicted in Fig. 1, a single blank in a sentence can have multiple possible words (i.e., "France", "Quebec" and "Cameroon") that fit the context of the original sentence. When used for nationality identification, it becomes imperative to supply a list of regions where French is spoken as an official language, ranked in order of population size.

Similarly, there may exist multiple plausible curves for patching an MDS. The *top-1* method depicted in (11) selects the candidate with the highest probability for each missing data point. If the PDF of the missing data has a higher peak or are more narrowly concentrated around a particular value (see the red PDF in Fig. 5), this method may have a higher accuracy in selecting the best candidate. However, when the PDF curve is a somewhat flattened one (see the green distribution in Fig. 5), selecting only the best candidate will greatly limit the inclusion of the original missing data point.

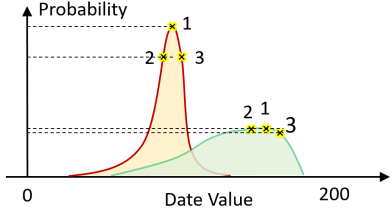

Fig. 5. Illustration of the probability distribution in $D$.

Therefore, we extend the top-1 method to the top-2 method, by comparing two selection approaches. Note that this approach can be readily extended to top-X candidate selection by repeatedly applying the same selection criterion.

**Method 1**: The most straightforward method entails choosing the candidate with the second-highest probability from each probability distribution to generate a secondary set of profiles. As depicted in Fig. 6, the continuous blue curve connecting the blue circles represents the curve produced by the top-1 candidates, whereas the dashed blue curve connecting the blue triangles represents the patching curve generated by directly selecting the top-2 candidates using Method 1.

Nonetheless, Method 1 has a major drawback - it disregards the autocorrelation among adjacent data points. In practice, opting for the second candidate may cause the PDFs of subsequent missing data points to shift accordingly.

**Method 2**: To overcome the inherent limitation of Method 1, we introduce an iterative selection method centered on the initial identification of "fork points". By using the identified fork points as reference pivots, we can create the top-2 candidate curve accounting for the autocorrelation among the subsequent data points.

Define the "right-side" (or "left-side") fork point as the first point, counting from the rightmost (or leftmost) side of the MSD, where the probability difference between the top-1 and top-2 candidates is less than $e$. As depicted in Fig. 6, after identifying the fork points, we can start an iterative process to estimate the top-1 values for the remaining missing data points, one at a time, by shifting the load profile either forward or backward. This method ensures that the shifted load profile maintains the same masking position as the original one, effectively capturing the shift in PDFs for subsequent data points.

The resultant top-2 curve is shown by the red continues line in Fig. 5 and the detailed algorithm description is depicted in Algorithm 1.

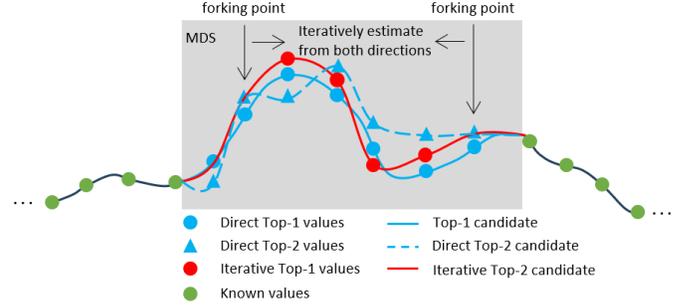

Fig. 6. Illustration of the top-2 candidate selection process.

---

**Algorithm 1** *Iterative Top Candidates Selection.*

Given the output of BERT-PIN $D$, find the fork points located on both the left and right sides of the top-1 curve. Next, replace the top-1 values at the fork points with the top-2 values. Then, iteratively generate the remaining missing data using top-1 values. Note that $e$ is the threshold for selecting the fork points.

---

Let $l = (t_{end} - t_{start})/2$
**for** $t = t_{start} : t_{start} + l$ **do**
    # find the left fork point.
    **if** $top1(D_t) - top2(D_t) < e$ **do**
        $fork_{left} = t$
        $x_t^{masked} = $ index of $top2(D_t)$
        **break**
    **else**
        $x_t^{masked} = $ index of $top1(D_t)$
**end for**

**for** $t = t_{end} : t_{end} - l$ **do**
    # find the right fork point.
    **if** $top1(D_t) - top2(D_t) < e$ **do**
        $fork_{right} = t$
        $x_t^{masked} = $ index of $top2(D_t)$
        **Break**
    **else**
        $x_t^{masked} = $ index of $top1(D_t)$
**end for**

**for** $k = fork_{left} - t_{start} : l$ **do**
- shift the daily profile by $k$ steps to the left.
- feed shifted data into the BERT-PIN model.
    $D' = BERT(X^{shifted}, T^{shifted})$
- update the first unknown value in $X^{masked}$
    $x_{t_{start}+k}^{masked} = $ index of $top1(D'_{t_{start}})$
**end for**

**for** $k = t_{end} - fork_{right} : l$ **do**
- shift the daily profile by $k$ steps to the right.
- feed shifted data into the BERT-PIN model.
    $D' = BERT(X^{shifted}, T^{shifted})$
- update the last unknown value in $X^{masked}$
    $x_{t_{end}-k}^{masked} = $ index of $top1(D'_{t_{end}})$
**end for**

Use the final $X^{masked}$ as the restored *top-2* load profile, $\hat{X}$.

---

### E. Performance Metrics

The performance metrics used for evaluating the accuracy of the restored data segments are listed in Table I. They are

calculated as

$$MPE = \frac{1}{K} \sum_{t=1}^{K} \frac{|\hat{x}_t^m - x_t^m|}{x_t^{event}} \qquad (12)$$

$$RMSE = \sqrt{\frac{1}{K} \sum_{t=1}^{K} (\hat{x}_t^m - x_t^m)^2} \qquad (13)$$

$$PKE = \frac{|\hat{x}^{MAX} - x^{MAX}|}{x^{MAX}} \qquad (14)$$

$$VLE = \frac{|\hat{x}^{MIN} - x^{MIN}|}{x^{MIN}} \qquad (15)$$

$$EGYE = \frac{|\sum_{t=1}^{K} \hat{x}_t^m - \sum_{t=1}^{K} x_t^m|}{\sum_{t=1}^{K} x_t^{event}} \qquad (16)$$

$$FCE = \frac{1}{K} \sum_{t=1}^{K} \frac{|FFT(\hat{x}_t^m) - FFT(x_t^m)|}{FFT(x_t^m)} \qquad (17)$$

where $K$ is the total number of data points in the MDS, $\hat{x}$ represents the restored data segment, $x^{MAX}$ and $x^{MIN}$ are the maximum and minimum power values in the MDS, respectively, and $FFT$ stands for Fast Fourier Transform. These indices offer insights into different aspects of errors, including point-to-point errors, inaccuracies in peak and valley points, discrepancies in total energy consumption, and errors within the frequency domain components.

TABLE I
ACCURACY EVALUATION METRICS

| No. | Indexes |
|---|---|
| 1 | Mean Percentage Error (MPE) |
| 2 | Root Mean Square Error (RMSE) |
| 3 | Peak Error (PKE) |
| 4 | Valley Error (VLE) |
| 5 | Energy Error (EGYE) |
| 6 | Frequency Components Error (FCE) |

## III. SIMULATION RESULTS

In this section, we first evaluate the impact of masking strategies on BERT-PIN performance. We then assess the ability for BERT-PIN to recover varying numbers of MDSs and compare it with three leading methods: LSTM [12], SAE [13], and Load-PIN [1]. Finally, we compare the performance of the top-1 and top-2 candidate methods for estimating the baseline in a CVR event.

### A. Data Preparation

The load profiles used in this study consists of 15-minute resolution smart meter data obtained from 8,000 users in North Carolina between 2018 and 2020. The corresponding temperature data is downloaded from the National Oceanic and Atmospheric Administration (NOAA) [32] website and is used as a second modality input.

We randomly select 2,000 users from the pool of 8,000 and combine their data to form feeder-level load profiles. These aggregated load profiles are aligned with the temperature profile and then normalized based on the annual load and temperature peaks, respectively. Then, we partition the profiles into either daily (96 data points) or weekly profiles (672 data points). Each missing load data segment is consistently set at 4 hours (16 data points), a choice guided by the observation that around 70% of missing load data segments in actual utility data are less than 4 hours in duration. It's important to note that there are no missing data segments in the temperature profile. The dataset is divided into an 80% training set and a 20% testing set.

The hyperparameters of the BERT-PIN model are summarized in Table II.

TABLE II
HYPERPARAMETERS SELECTED FOR THE BERT-PIN MODEL

| Parameter | Values |
|---|---|
| Learning rate | 1e-4 |
| Local loss weight $\lambda$ | 0.8 |
| Batch size | 16 |
| Training epochs | 100 |
| Number of heads | 2 |
| Number of transformer layers $\ell$ | 2 |

### B. Case 1: Restoration of One MDS using Central Masking

In Case 1, our objective is to evaluate the performance of BERT-PIN when restoring a single 4-hour MDS. In the training process, we randomly apply a 4-hour mask to the 3-load profile. Subsequently, we extract a 10-hour data segment both preceding and following the masked data section, creating a 24-hour load profile. This resulting daily load profile, with a 4-hour gap in its center, is used as input for the BERT-PIN model. Given that the MDS is consistently positioned at the center of the load profile, we refer to this method as the "central masking approach."

Several restored daily load profiles are shown in Fig. 7, while the corresponding performance metrics are depicted in Fig. 8 and summarized in Table III. As illustrated in Fig. 7, the data segments restored by BERT-PIN (highlighted in red) exhibit a remarkable proximity to the ground truth (indicated by the green lines). Following BERT-PIN, the GAN-based LOAD-PIN method is the second-best performer, with LSTM and SAE demonstrating comparable performance. This is further confirmed by the error distribution shown in Fig. 8, where BERT-PIN (red) exhibits the smallest medians and ranges of errors, highlighting its superior performance.

Table III shows that BERT-PIN outperforms the best benchmark method by margins ranging from 3% to 21% in five of the six metrics. It's worth noting that BERT-PIN exhibits only a slightly higher FCE compared to Load-PIN. Nevertheless, these findings demonstrate that the proposed model excels in accurately restoring the missing data in load profile inpainting problems, in the case of a single MDS using central masking.

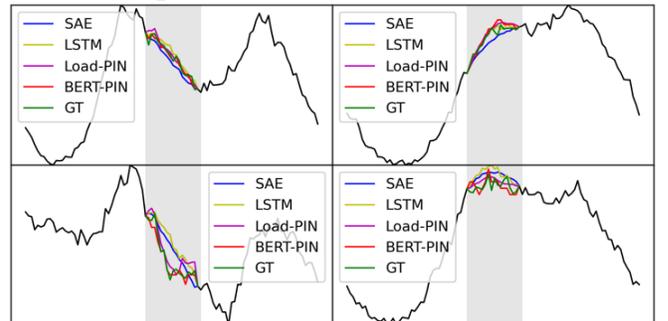

Fig. 7. Examples of missing data restoration with central-mask using different models: SAE (blue), LSTM (yellow), Load-PIN (magenta), BERT-PIN (red), and ground truth (green).

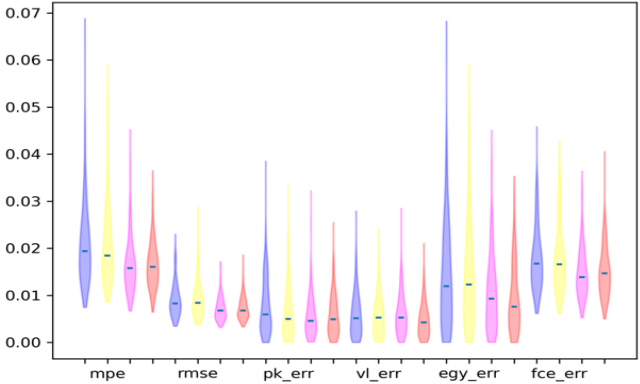

Fig. 8 Error distributions of different models: SAE (blue), LSTM (yellow), Load-PIN (magenta), BERT-PIN (red).

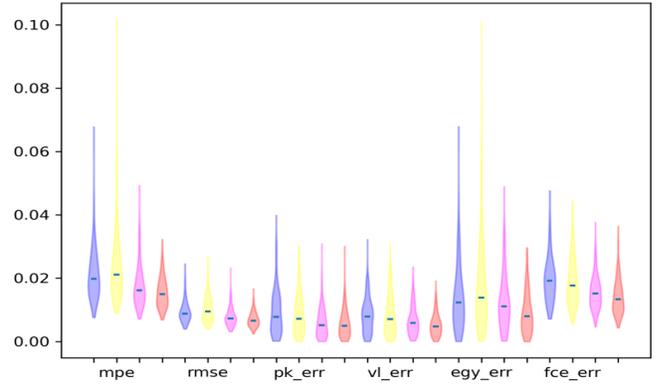

Fig. 10. Error distributions of different models: SAE (blue), LSTM (yellow), Load-PIN (magenta), BERT-PIN (red).

TABLE III
ERRORS OF SINGLE EVENT CENTRAL MASKING CASE

|  | SAE (%) | LSTM (%) | Load-PIN (%) | **BERT-PIN (%)** | Improvement |
|---|---|---|---|---|---|
| MPE | 2.089 | 2.144 | 1.683 | **1.612** | 3.74% |
| RMSE | 0.8882 | 0.9032 | 0.9067 | **0.6992** | 21.27% |
| PKE | 0.7443 | 0.6281 | 0.5731 | **0.5438** | 5.11% |
| VLE | 0.6235 | 0.5774 | 0.5764 | **0.4815** | 16.46% |
| EGYE | 1.440 | 1.516 | 1.059 | **0.8879** | 16.16% |
| FCE | 1.781 | 1.743 | 1.452 | **1.467** | -1.03 |

TABLE IV
ERRORS OF SINGLE EVENT PEAK MASKING CASE

|  | SAE (%) | LSTM (%) | Load-PIN (%) | **BERT-PIN (%)** | Improvement |
|---|---|---|---|---|---|
| MPE | 2.231 | 2.414 | 1.670 | **1.523** | 8.80% |
| RMSE | 1.035 | 1.112 | 0.7951 | **0.7404** | 6.88% |
| PKE | 1.065 | 0.8491 | 0.6183 | **0.5130** | 17.10% |
| VLE | 0.8687 | 0.8991 | 0.6185 | **0.5870** | 5.09% |
| EGYE | 1.525 | 1.762 | 1.165 | **0.8410** | 27.81% |
| FCE | 2.138 | 2.046 | 1.615 | **1.509** | 6.56% |

## C. Case 2: Restoration of One MDS using Peak Masking

In real-time operation, an MDS may not occur at the center of a daily profile. For example, when performing CVR for peak shaving, the baseline load consumption during the CVR period can be viewed as an MSD. Such MSDs occur during the peak load hours. To derive the CVR baseline, we train the BERT-PIN by only masking the 4-hour load peak periods on each day.

Several restored daily load profiles are shown in Fig. 9, while the corresponding performance metrics are depicted in Fig. 10 and summarized in Table IV. It is evident from the outcomes that BERT-PIN (represented by the red lines) exhibits the smallest medians and ranges in this case. Furthermore, Table IV provides the mean errors for each distribution shown in Fig. 9. The results clearly indicate that BERT-PIN achieves the lowest errors and outperforms the benchmark methods by 5%-27% across all six indexes. This demonstrates that our model consistently outperforms the existing approaches across various masking scenarios.

## D. Top-2 Candidate Selection

As described in section II, BERT-PIN has the capability to generate multiple restored load profiles using *Top Candidate Selection* methods. In this section, we compare the two Selection methods discussed in section II.D, taking the selection of the top-2 candidates as an illustrative example. It's important to note that this method can be employed iteratively to choose the top-X candidates.

The results obtained from central masking and peak masking are presented in Fig. 11 and Fig. 12, respectively. BERT-PIN is the default *top-1* candidate where the candidate with the highest probability value is selected. BERT-PIN_2 results are obtained using Method 1, where candidates having the *second highest* probability is selected. BERT-PIN_2i results are obtained using method 2, where an iterative method is used to select a fork point, based on which, subsequent restoration candidates are selected.

The errors of *top-1* candidate and *top-2* candidate with different parameters are computed and presented in Table V.

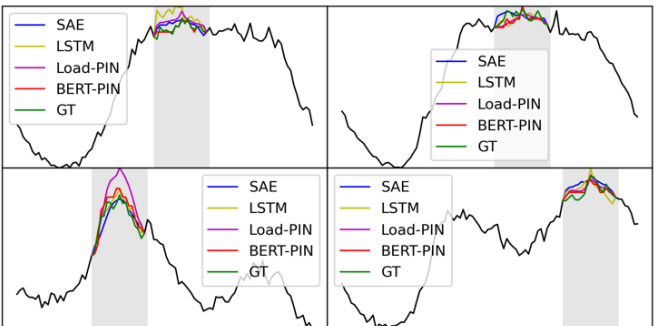

Fig. 9. Examples of missing data restoration with central-mask using different models: SAE (blue), LSTM (yellow), Load-PIN (magenta), BERT-PIN (red), and ground truth (green).

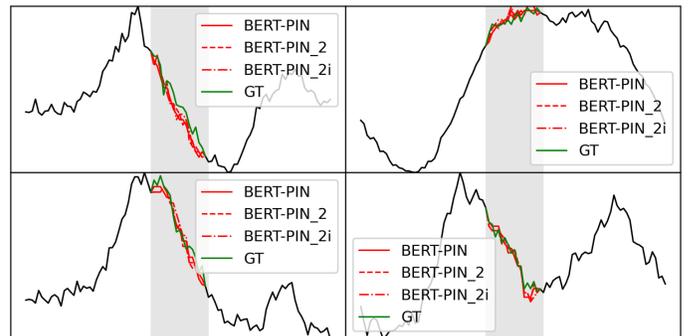

Fig. 11. Examples of top-1 and top-2 restored MDSs with central-mask.

TABLE V
ERRORS OF TOP-1 CANDIDATE, TOP-2 CANDIDATES AND COMBINED OUTPUTS (%)

| | | BERT-PIN | BERT-PIN_2 | BERT-PIN_2i | | | | | | Combine |
| | | | | e=0.8 | e=0.5 | e=0.3 | e=0.1 | e=0.05 | e=0.02 | |
|---|---|---|---|---|---|---|---|---|---|---|
| Central mask | MPE | **1.612** | 1.779 | 2.803 | 2.526 | 2.435 | 2.157 | 1.88 | **1.745** | 1.274 |
| | RMSE | **0.6992** | 0.7967 | 1.402 | 1.206 | 1.091 | 1.01 | 0.829 | **0.769** | 0.566 |
| | PKE | **0.5438** | 0.5617 | 0.825 | 0.775 | 0.698 | 0.642 | 0.559 | **0.575** | 0.477 |
| | VLE | **0.4815** | 0.4572 | 0.979 | 0.69 | 0.559 | 0.645 | 0.472 | **0.453** | 0.336 |
| | EGYE | **0.8879** | 0.9905 | 1.795 | 1.635 | 1.536 | 1.324 | 1.099 | **1.013** | 0.702 |
| | FCE | **1.467** | 1.638 | 3.006 | 2.53 | 2.194 | 2.188 | 1.74 | **1.609** | 1.218 |
| | PoCP | - | 45.02% | 24.88% | **26.95%** | 24.18% | 16.18% | 11.10% | **4.97%** | - |
| Peak mask | MPE | **1.523** | **1.744** | 2.556 | 2.433 | 2.407 | 2.138 | 1.87 | 1.761 | 1.211 |
| | RMSE | **0.7404** | 0.9144 | 1.317 | 1.173 | 1.211 | 1.071 | 0.896 | **0.899** | 0.577 |
| | PKE | **0.5130** | **0.5917** | 1.044 | 0.939 | 0.927 | 0.871 | 0.665 | 0.663 | 0.426 |
| | VLE | **0.5870** | 0.9260 | 0.827 | 0.617 | 0.817 | 0.573 | **0.543** | 0.669 | 0.407 |
| | EGYE | **0.8410** | 0.9582 | 1.618 | 1.441 | 1.412 | 1.301 | 1.043 | **0.986** | 0.633 |
| | FCE | **1.509** | 1.942 | 2.447 | 2.18 | 2.273 | 2.007 | **1.727** | 1.78 | 1.209 |
| | PoCP | - | 45.12% | 23.88% | **24.40%** | 22.94% | 17.29% | 12.25% | **6.29%** | - |

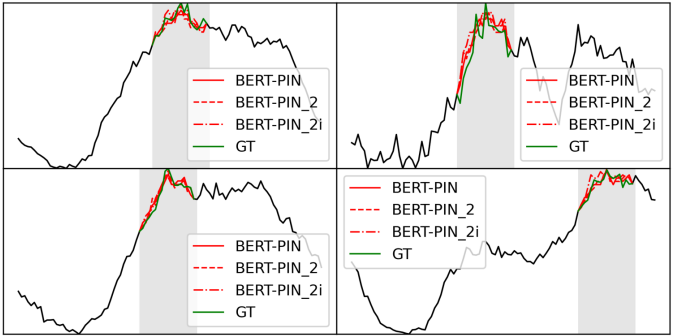

Fig. 12. Examples of top-1 and top-2 restored MDSs with peak masking.

Given that the process can be repeated for generating an ensemble of candidates for MDS restoration, we proceed to assess the quality of the top-2 results by calculating the "percentage-of-closer-points" (PoCP). This metric signifies the percentage of estimated points that are closer to the ground truth when compared to those estimated using the top-1 method, indicating the potential expansion of inclusion through the incorporation of the top-2 restored MDS.

From the results shown in Table V, we made the following observations:

- As expected, the result accuracy of the *top-1* method is higher than the *top-2* method.
- Nevertheless, the *top-2* method can generate points that are closer to the ground truth. This is because the BERT output is generated based on probabilities, and the second-best candidate could also be close to the actual missing data, just with reduced likelihood.
- Compared with the direct *top-2* candidate selection method, iterative *top-2* candidate selection method has lower PoCP. However, since the points selected independently lack autocorrelation among adjacent points, we consider the results to be less valuable for time-series MDS.
- Among all iterative top-2 methods, when *e* = 0.5, the PoCP is the highest. This demonstrate that the fork point selection influences the range of inclusion. When *e* becomes smaller, the iterative top candidates selection algorithm tends to identify a fork point later. Consequently, the top-2 candidate curve closely resembles the top-1 candidate, leading to identification of fewer "better" points.

- When we merge the top-1 and top-2 candidate by selecting the best value for each timestamp (column highlighted in yellow), we demonstrate the benefit of incorporating both candidates when performing downstream tasks. This advantage is reflected in the wider range of data encompassed because, with both curves, we capture data points closer to the original missing data points.

### E. Restoration of Multiple MDSs

In this section, we showcase the performance of BERT-PIN for restoring multiple MDSs within a weekly load profile. This use case is essential for Conservation Voltage Reduction (CVR) baseline estimation. CVR is a frequently adopted strategy among utility companies to manage peak loads. During CVR events, the voltage on a distribution feeder is deliberately decreased by 2-4% over a period ranging from 1 to 4 hours. CVR can be executed on multiple days in a hot summer week or a cold winter week. As depicted in Fig. 13, accurately assessing the actual load reduction achieved through CVR often involves estimating the baseline energy consumption (indicated by the red lines) that would occur in the absence of CVR implementation.

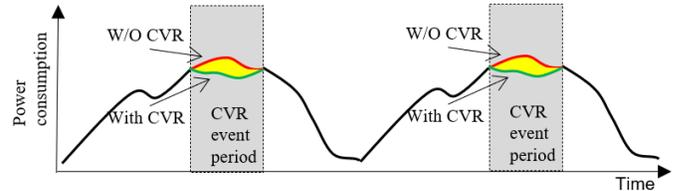

Fig. 13. An illustration of the CVR baseline estimation

To train BERT-PIN for CVR baseline estimation, we apply 4-hour masks targeting the peak load periods within weekly load profiles. The number of masks is randomly select between 1 and 7, thereby representing different numbers of CVR event days. The hyperparameters of the BERT-PIN model remain consistent with those employed in the single-event scenario. As can be seen from Fig. 14 and Table VI, BERT-PIN outperform all existing methods by a fairly large margin.

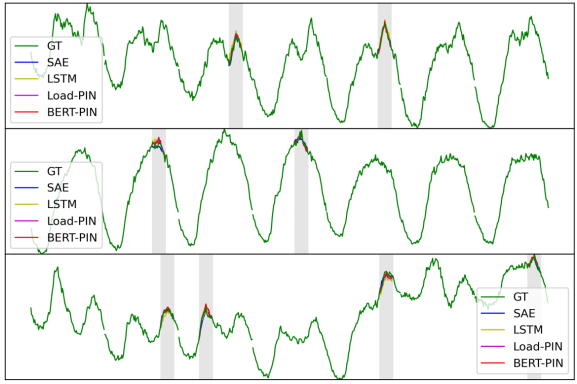

Fig. 14. Missing data restoration examples: SAE (blue), LSTM (yellow), Load-PIN (magenta), BERT-PIN (red), and ground truth (green).

TABLE VI
ERRORS OF MULTIPLE EVENTS PEAK MASKING CASE

|  | SAE (%) | LSTM (%) | Load-PIN (%) | BERT-PIN (%) | Improvement |
|---|---|---|---|---|---|
| MPE | 7.615 | 6.389 | 5.168 | **4.837** | 6.40% |
| RMSE | 3.310 | 2.848 | 2.301 | **2.221** | 3.48% |
| PKE | 3.362 | 2.193 | 1.769 | **1.746** | 1.30% |
| VLE | 2.697 | 2.407 | 1.691 | **1.627** | 3.78% |
| EGYE | 4.874 | 3.822 | 3.257 | **2.699** | 17.13% |
| FCE | 6.828 | 5.601 | 4.808 | **4.501** | 6.39% |

Next, we conduct a comparative analysis of two BERT-PIN models in their role of restoring the CVR baseline: *BERT-PIN-Day* and *BERT-PIN-Week*. *BERT-PIN-Day* uses the 24-hour load profile specific to each CVR day as input, allowing it to recover the CVR baseline for that particular day. In contrast, *BERT-PIN-Week* uses the weekly load profile as input, enabling it to simultaneously recover multiple CVR baselines within the week.

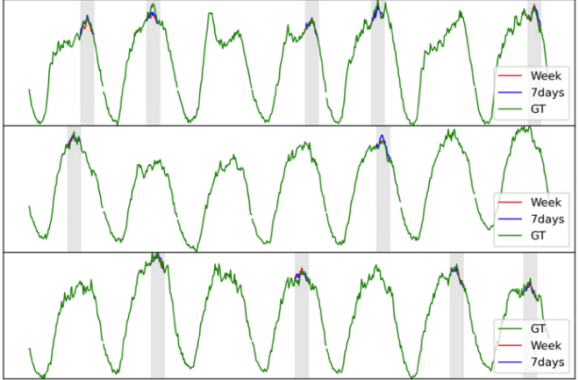

Fig. 15. Comparing BERT-PIN-Week (red) and BERT-PIN-Day (blue).

TABLE VII
Errors of BERT-PIN-Week and BERT-PIN-Day

| Performance Metrics | **BERT-PIN-Week** | BERT-PIN-Day |
|---|---|---|
| MPE | **4.837%** | 5.244% |
| RMSE | **2.221%** | 2.447% |
| PKE | **1.746%** | 1.805% |
| VLE | **1.627%** | 2.017% |
| EGYE | **2.699%** | 3.027% |
| FCE | **4.501%** | 4.955% |
| Size of training data | 310k weekly loads | 230k daily loads |
| Model training time | 20 hours | 8 hours |

For instance, if there are three CVR event days within a hot summer week, *BERT-PIN-Day* processes the 24-hour load profile for each CVR day individually to recover the respective CVR baselines one by one. On the other hand, BERT-PIN-Week employs the entire weekly load profile to recover all three CVR baselines concurrently.

From results shown in Fig. 15 and Table VII, we conclude that *BERT-PIN-Week* demonstrates superior performance compared to *BERT-PIN-Day*, with smaller errors. This is because *BERT-PIN-Week* uses weekly load profiles as input, enabling the attention mechanism to gain a deeper understanding of the load patterns and resulting in a more precise estimation of the missing data. However, although BERT-PIN-Week outperforms *BERT-PIN-Day* in terms of accuracy, but it requires a larger amount of training data and substantial longer training time. This shows a trade-off between computing costs and performance improvements.

## IV. CONCLUSION

In this study, we introduced an innovative load profile inpainting framework, BERT-PIN, which capitalizes on the capabilities of the BERT model to restore multiple missing data segments. BERT-PIN demonstrated superior performance in both single-MDS and multiple-MDS recovery, surpassing existing methods. Furthermore, we demonstrated its practical application in the domain of demand response baseline estimation, using CVR baseline estimation as a representative example. Furthermore, we introduce top-X candidate selection methods, demonstrating that although the top-1 candidate maintained the highest accuracy, the inclusion of top-2 candidates expanded the coverage of estimated missing data, encompassing a broader range of recovered data points. This underscores the potential of using BERT-PIN as an ensemble generation tool, facilitating more comprehensive analyses during periods with missing data.

Our future research endeavors will extend the utility of the BERT model as a versatile, pre-trained tool for various downstream tasks, including super resolution and load disaggregation.